\documentclass[12pt]{article}
\usepackage{graphicx,amsmath,amssymb}
\usepackage{enumerate}
\usepackage{psfrag}
\usepackage[usenames,dvipsnames]{xcolor}

\usepackage{color}   

\usepackage{cite}
\let\oldbibliography\thebibliography
\renewcommand{\thebibliography}[1]{%
  \oldbibliography{#1}%
  \setlength{\itemsep}{-1pt}%
}

\newcommand{\EQ}{\begin{equation}}
\newcommand{\EE}{\end{equation}}
\newcommand{\EQA}{\begin{eqnarray}}
\newcommand{\EEA}{\end{eqnarray}}

\usepackage{ulem}
\definecolor{purple}{rgb}{0.459,0.109,0.538}

\begin{document}

\title{\vspace*{-2cm} 
Epidemiological and evolutionary analysis of the 2014 Ebola virus outbreak}
\author{Marta \L uksza${}^{\, \rm a,b}$, Trevor Bedford${}^{\, \rm c}$, and Michael L\"assig${}^{\, \rm a, *}$ }

\date{\small
(a)  Institute for Theoretical Physics, University of Cologne, Z\"ulpicherstr.~77, \\ 50937 K\"oln,
Germany \\
(b) Institute for Advanced Study, 1 Einstein Dr, Princeton, NJ 08540, USA 
\\
(c) Vaccine and Infectious Disease Division, Fred Hutchinson Cancer Research Center, 1100 Fairview Ave N, Seattle, WA 98109, USA
\\
* To whom correspondence should be addressed. Email: lassig@thp.uni-koeln.de
}
\maketitle

\noindent 
{\bf 
The 2014 epidemic of the Ebola virus is governed by a genetically diverse viral population. In the early Sierra Leone outbreak, a recent study has identified new mutations that generate genetically distinct sequence clades~\cite{Gire14}. Here we 
find evidence that major Sierra Leone clades have systematic differences in growth rate and reproduction number. If this growth heterogeneity remains stable, it will generate major shifts in clade frequencies and influence the overall epidemic dynamics on time scales within the current outbreak. Our method is based on simple summary statistics of clade growth, which can be inferred from genealogical trees with an underlying clade-specific birth-death model of the infection dynamics. This method can be used to perform realtime tracking of an evolving epidemic and identify emerging clades of epidemiological or evolutionary significance.
}
\bigskip 

In a recent study, Gire et al.~present a comprehensive genomic analysis of the current Ebola virus epidemic in its early growth phase~\cite{Gire14}. Based on a near-complete sequence sample of cases, they reconstruct the seeding event and the subsequent rapid expansion of the epidemic in Sierra Leone in May and June 2014. Estimates of the basic reproduction number $R_0$ for Sierra Leone vary between $1.7$ and 2.2~\cite{WHO14, Stadler14}, the most recent estimate of the reproduction number  is $R = 1.4 \pm 0.1$~\cite{WHO14}. In the current outbreak, the Ebola virus has evolved new genetic diversity. In particular, Gire et al.~identify early mutations that delineate major sequence clades in Sierra Leone~\cite{Gire14}.

In this paper, we develop a method for the statistical analysis of epidemic processes that is based on simple summary statistics of individual clades. Using this method, we study evolutionary consequences of the genetic structure in the Sierra Leone population.
We identify a sequence clade that has larger growth rate and reproduction number than its genetic background. We discuss likely causes for growth heterogeneity within the Sierra Leone outbreak, which include transient epidemiological factors and adaptive evolution.

\subsubsection*{Genealogical trees of the Sierra Leone outbreak} 

We construct a Bayesian posterior ensemble of genealogical trees from the Sierra Leone sequences of ref.~\cite{Gire14} (see Methods). Internal nodes represent new infections and external nodes mark the termination of infectivity in an individual; however, this type of analysis does not attempt to reconstruct detailed transmission chains \cite{Ypma13, Didelot14}. 

The sample tree of Fig.~1 shows three major sequence clades, in accordance with the results Gire et al. (see Fig.~3b and Fig.~4a of ref.~\cite{Gire14}). Four point mutations on an early branch separate the ancestral clade~1 from clade~2, which carries the derived alleles and covers 36\% of the Sierra Leone sequences. These mutations probably occurred prior to the seeding of the epidemic in Sierra Leone~\cite{Gire14}. A further intergenic mutation between genes VP30 and VP24 at position 10218 occurred within Sierra Leone. This mutation separates clade~2 from clade~3, which carries the derived allele and covers 56\% of the sequences. VP30 is a virus-specific transcription factor that mediates transcription activation \cite{Martinez08}, while VP24 suppresses interferon production \cite{Leung06} and also inhibits viral transcription and replication \cite{Watanabe07}. 
All other mutations occur on peripheral branches of the tree and have derived allele frequencies below 8\%.

\subsubsection*{Inference of clade-specific growth} 

From these trees, we record the growth dynamics of each clade until a given time $t$ by three observables: the cumulative number of new infections $I_\alpha (t)$, the cumulative number of terminations of infectivity $E_\alpha (t)$, 
and the cumulative infectivity period 
on all clade lineages $T_\alpha (t)$, where $\alpha$ denotes clades 1,2,3. Fig.~2a shows these observables averaged over 10 high-scoring trees; data from individual trees show only minor deviations from these averages (Fig.~S2). For the derived clades, the number of internal nodes, the number of external nodes and total tree length increase more rapidly in clade 3 than in clade 2 (Fig.~2a). Hence, these data suggest an exponential growth with clade-specific growth rates. 
The ancestral clade~1 is clearly subleading and probably extinct after the end of May (including these strains into a combined clade~1\&2 does not affect the results of the following analysis). 

To determine clade growth rates and their statistical significance, a probabilistic phylodynamic model is needed~\cite{Volz13}. We use a birth-death model: new infections and terminations of infectivity are treated as independent Poisson processes on each lineage, which are characterized by clade-specific rates $b_\alpha$ and $d_\alpha$. The applicability of this model to the Ebola epidemic is supported by the statistics of successive infection and termination events in individual lineages, which occur at a clade-specific rate $(b_\alpha +d_\alpha)$ (Methods and Fig.~2b). The rates $b_\alpha$ and $d_\alpha$
determine the growth rate  $r_\alpha = b_\alpha - d_\alpha$ and the reproductive number $R_\alpha = b_\alpha / d_\alpha$ of each clade~\cite{Stadler10, Stadler12}. If we identify the growth rates $r_\alpha$ with Malthusian fitness, any such birth-death model can be interpreted as a fitness model. 

For a given sample of posterior trees, we can infer the underlying clade-specific birth-death model using a Bayesian approach that accounts for uncertainty in tree reconstruction~\cite{Stadler10, Stadler12}. Importantly, the likelihood of the model with rates $b_\alpha$ and $d_\alpha$ depends on the summary statistics $I_\alpha$, $E_\alpha$, and $T_\alpha$ of our sequence sample  in a simple way (Methods, Fig.~S3).
We evaluate this likelihood for a family of models with 
clade~2 and clade~3 growing at distinct rates $r_2$ and $r_3$, and we denote their growth rate difference by $s = r_3 - r_2$. 
We also evaluate the likelihood of background models that have common rates $b_2 = b_3$ and $d_2 = d_3$, which implies $s = 0$. 
Our analysis reveals a substantial growth difference between clades: in the global maximum-likelihood model, clade~3 has growth rate $r^*_3 = 0.104$,
compared to $r_2^* = 0.051$ 
for clade~2. The difference $s^* = 0.053$ is of the same order of magnitude as the absolute growth rates (all rates are measured in units of 1/day).  
The best background model has 
a growth rate $r_0 = 0.082$ 
for the combined clade 2\&3, which is consistent with the uniform growth parameters inferred in ref.~\cite{Stadler14}.
The log-likelihood difference between these models is $4.74$, which implies $P = 8.7 \times 10^{-3}$ (log-likelihood ratio test).

\subsubsection*{Epidemiological determinants of clade-specific growth} 

The growth difference between early clades may be due to transient epidemiological effects, such as superspreading~\cite{Gire14,Volz14b}, or be associated with environmental factors. Individual superspreading events may give a short-time growth rate difference between small clades, but multiple events average out over longer time scales and do not affect systematic growth rate differences between genetic clades. Given the uncertainties in reconstructing individual transmission chains, the evidence for multiple superspreading events in the Sierra Leone outbreak remains limited~\cite{Stadler14}. 

Similarly, we expect that environmental effects are more pronounced for the absolute growth rates $r_2$ and $r_3$, while their difference $s$ remains more stable. Because the data set of this study comes from a confined geographical region in eastern Sierra Leone (most later cases occurred in the Jawie chiefdom)~\cite{Gire14}, geographical heterogeneity is likely to have limited effects on the inferred clade dynamics over the period of the data set. 
However, growth rates may change if any of the Sierra Leone clades establishes itself beyond its initial geographical region.

\subsubsection*{Adaptive evolution within an epidemic}

If the observed growth heterogeneity remains stable, we can use clade-specific evolutionary models to predict the future evolution of clade frequencies~\cite{Luksza14}. This analysis predicts that clade~3 will outgrow clade~2 within a period of order $1/s^*$, which amounts to a few months (counted from the end of June 2014, Methods). Evolutionary clade displacement is expected, in particular, if the growth difference between clades~2 and~3 is caused by their genetic difference. In that case, the mutation that separates these clades conveys a fitness increase $s^* = 0.055$, and the resulting changes in clade frequencies are adaptive. 

Adaptive evolution within an epidemic has two important characteristics. First, clade frequency shifts caused by fitness differences are predicted to occur even when increased immunity and intervention curb the overall number of infections. Second, adaptive evolution feeds back on the overall dynamics of infections: it increases the mean population fitness, which accelerates the epidemic and increases its peak size (for details, see Methods). 

Forthcoming genetic data will allow a clade-specific inference on broader regional and temporal scales, which is essential to distinguish between epidemiological and evolutionary determinants of clade dynamics. If growth differences are primarily caused by external factors, there are only weak correlations between growth rates and genetic differences. In the case of adaptive evolution, we  expect a continual turnover of clades that is caused by their genetic differences: high-fitness alleles tend to fixation, while further new alleles gain substantial population frequencies and add fitness variance to the population.

\subsubsection*{Conclusion} 

We have developed a method to infer clade-resolved epidemiological parameters from timed sequence data of an evolving epidemic. Our method uses only simple phylodynamic summary statistics that measure the growth dynamics of individual clades and are robust under fluctuations of inferred pathogen genealogies. We use these summary statistics to infer the rates $b_\alpha, d_\alpha$ of a clade-specific birth-death infection model, which distinguishes our method from related inference schemes based on coalescent trees~\cite{Luksza14, Neher14}. The underlying Bayesian statistical analysis distinguishes systematic shifts in clade frequencies from stochastic effects in the transmission process that contribute to genetic drift. The same effects also generate large fluctuations in the overall number of infections during the initial ``stuttering'' phase of an epidemic \cite{LloydSmith09}. 

In the 2014 Ebola epidemic, we have identified a genetic variant 
that has a substantially higher growth rate than its progenitor clade. We conclude that a viral epidemic can develop strong growth heterogeneity even on the limited temporal and spatial scales of its initial outbreak. If that heterogeneity has a genetic cause, our analysis suggests that selection can shape a fast-evolving pathogen on the time scales of a single epidemic. This mode of adaptation is much faster than the adaptive changes between epidemics addressed in previous studies~\cite{Gire14, Volz14b}. 
However, we caution against over-interpretation of our early results. All predictions starting from the initial phase of an epidemic are probabilistic extrapolations; they are based on limited data and subject to confounding factors such as variation in sampling density. 
As more sequence data emerge, updated clade-specific inference will suggest targets for detailed epidemiological investigation and 
provide predictive insight into the dynamics of the epidemic. This joint evolutionary and epidemiological analysis will suggest targets for detailed epidemiological investigation and provide predictive insight into the dynamics of the epidemic.

\subsubsection*{Materials and Methods}

\paragraph{Sequence data and tree reconstruction.}
We use the Ebola virus sequence data of Gire at al. \cite{Gire14}. These data consist of complete genomes isolated from 78 Sierra Leone patients between late May and mid June 2014, together with three Guinea sequences from March 2014 used as outgroup. 
These data cover more than 70\% of all cases reported until June 16, 2014~\cite{Gire14}. 

Our analysis is based on consensus sequences for each patient. In addition, deep-sequencing identifies sequence diversity within patients~\cite{Gire14}. Sequence site 10218, which separates clade~2 from clade~3, is polymorphic in 12 of 78 individuals, with minor allele frequencies $< 25\%$ in 11 of 12 cases. Therefore, the consensus sequences provide an accurate representation of the allele frequencies in the viral population. We reconstruct genealogical trees from these sequences using the BEAST software package~\cite{BEAST17}. The inference is based on a demographic model with exponential growth, a strict molecular clock and an HKY nucleotide substitution model \cite{Hasegawa85}.

\paragraph{Clade-specific birth-death models.}
We analyze the tree data using birth-death models, because our sequence sample covers a large proportion of the underlying viral population. Simpler coalescent models require a small sampling fraction and are not applicable to this data set~\cite{Volz13, Volz14}.   
We consider birth-death models for a population that consists of $k$ sequence clades labelled by an index $\alpha = 1, \dots, k$ (with $k = 3$ for the data set of this study). Any such model is specified by $2k$ independent, stationary Poisson processes with rates $b_\alpha$ and $d_\alpha$, which govern birth (infection) and death (termination of infectivity) for the lineages in each clade. This model generates genealogical trees with a clade-specific distribution of branch lengths, which is given by a Poissonian with average $(b_\alpha + d_\alpha)$.

\paragraph{Tree summary statistics.}
For a given tree, we obtain summary statistics $I_\alpha (t)$, $E_\alpha (t)$, and $T_\alpha (t)$ as the total number of internal nodes, external nodes, and the sum of branch lengths up to time $t$ within a given clade $\alpha$. The node numbers determine the number of lineages (tree branches) in clade $\alpha$ present at a given time, $L_\alpha (t) = I_\alpha (t)  - E_\alpha (t)$, which is related to the cumulative branch length by  
\EQ
\dot T_\alpha (t) 
= L_\alpha (t) 
\EE
with dots denoting derivatives with respect to $t$. In a birth-death process with rates $b_\alpha$, $d_\alpha$, the expectation values of these quantities in a completely sampled population follow the deterministic dynamical equations 
\EQA
\dot I_\alpha (t) & = & b_\alpha L_\alpha (t), 
\nonumber 
\\ 
\dot E_\alpha (t) & = & d_\alpha L_\alpha (t), 
\nonumber 
\\ 
\dot L_\alpha (t) & = & r_\alpha L_\alpha (t) 
\qquad \mbox{with $r_\alpha = b_\alpha - d_\alpha$.}
\EEA
These equations lead to exponential growth, 
\EQA
I_\alpha (t)  & =  & b_\alpha T_\alpha (t)  =  C_\alpha \, b_\alpha \exp (r_\alpha t),
\nonumber 
\\
E_\alpha (t) & =  & d_\alpha T_\alpha (t)  =  C_\alpha \, d_\alpha \exp (r_\alpha t), 
\nonumber 
\\
T_\alpha (t) & =  & C_\alpha \, \exp (r_\alpha t), 
\label{exp}
\EEA
where $C_\alpha$ is common constant. Importantly, these equations determine the growth rate and the relative amplitudes of all three observables in terms of the two parameters $b_\alpha$, $d_\alpha$. In Fig.~S3, we compare the analytical growth dynamics with numerical simulations.

\paragraph{Sampling effects.} 
In order to link these growth dynamics to genealogical data, we must take into account the incomplete sampling of infection cases. 
We first consider temporally homogeneous sampling of a fraction $\rho$ of the cases, which applies to our data set for the bulk of the observation period (see the discussion below). 
If the dynamics of the full population follows a stationary birth-death process with rates $b_\alpha$ and $d_\alpha$, the genealogy of sampled lineages is described by a similar birth-death process with observed rates $b_\alpha (\rho)$ and $d_\alpha (\rho)$. These rates are lower than the full-population rates, because the sampling removes a part of the internal and external nodes from the genealogical tree, but they maintain the growth rate $b_\alpha (\rho) - d_\alpha (\rho) = r_\alpha$. The observed birth rate takes the form $b_\alpha (\rho)  =  b_\alpha [1- G_\alpha (\rho)]$, where $G_\alpha (\rho)$ is the probability that a subclade emerging from an infection event is not sampled in any of its lineages. Denoting the time of the initial infection by $t_0$, this probability is given by 
\EQ
G_\alpha (\rho) = \int_{t_0}^\infty [(1-\rho) d_\alpha + b_\alpha G_\alpha^2 (\rho) ] \, {\rm e}^{-(b_\alpha + d_\alpha) (t - t_0)} \, {\rm d}t = 
\frac{(1-\rho) d_\alpha + b_\alpha G_\alpha^2 (\rho)}{b_\alpha + d_\alpha},
\EE
which determines the observed rates 
\EQA 
b_\alpha (\rho) & = & \frac{1}{2} \left ( b_\alpha - d_\alpha + \sqrt{(b_\alpha - d_\alpha)^2 + 4 \rho b_\alpha d_\alpha} \right ) ,
\nonumber
\\
d_\alpha (\rho) & = & \frac{1}{2} \left ( d_\alpha - b_\alpha + \sqrt{(b_\alpha - d_\alpha)^2 + 4 \rho b_\alpha d_\alpha} \right ).
\label{transf}
\EEA
These rates interpolate between the full-population rates
\EQ
b_\alpha (\rho \! = \! 1) = b_\alpha, 
\qquad 
d_\alpha (\rho \! = \! 1) = d_\alpha
\EE
and the limit of low bulk sampling,
\EQ
b_\alpha (\rho \! = \! 0) = r_\alpha, 
\qquad 
d_\alpha (\rho \! = \! 0) = 0,
\EE
in which the genealogy reduces to a coalescent tree. The observed summary statistics $I_\alpha (t; \rho)$, $E_\alpha  (t; \rho)$, and $T_\alpha (t; \rho)$ follow the dynamical equations 
\EQA
\dot I_\alpha (t; \rho) & = & b_\alpha (\rho)  \, L_\alpha (t; \rho)
\nonumber 
\\ 
\dot E_\alpha (t; \rho) & = & d_\alpha (\rho)  L_\alpha (t, \rho) = \rho d_\alpha  L_\alpha (t), 
\nonumber 
\\ 
\dot T_\alpha (t; \rho) & = & L_\alpha (t; \rho), 
\nonumber 
\\
\dot L_\alpha (t;\rho) & = & r_\alpha  L_\alpha (t,\rho), 
\label{dot_f_rho}
\EEA
which again lead to exponential growth
\EQA
I_\alpha (t; \rho)  & =  & b_\alpha T_\alpha (t; \rho)  =  C_\alpha  (\rho) \, b_\alpha(\rho) \exp (r_\alpha t),
\nonumber 
\\
E_\alpha (t; \rho) & =  & d_\alpha T_\alpha (t; \rho)  =  C_\alpha  (\rho) \, d_\alpha(\rho) \exp (r_\alpha t) , 
\nonumber 
\\
T_\alpha (t) & =  & C_\alpha  (\rho) \, \exp (r_\alpha t) 
\label{exp_rho}
\EEA
with $C_\alpha (\rho) =  C_\alpha d_\alpha \rho / d_\alpha (\rho)$. As shown by comparison with equation (\ref{exp}), these quantities grow exponentially at the same rate as in the full population, but with modified relative amplitudes. 

If the sampling of the population ends at a given time $t_f$, the observed infection and termination rates become explicitly time-dependent. The infection rate takes the form $b_\alpha (t; t_f, \rho) = b_\alpha [1 - G_\alpha((t; t_f, \rho)]$, where $G_\alpha (t; t_f, \rho)$ is the probability that a subclade starting with an infection event at time $t$ is not sampled in any of its lineages before $t_f$. This probability is the solution of the differential equation
\EQ
\frac{\partial}{\partial t} G_\alpha (t; t_f,\rho) =   (b_\alpha + d_\alpha) G_\alpha (t; t_f,\rho) - b_\alpha G^2_\alpha (t; t_f,\rho) - (1 - \rho) d_\alpha  
\EE 
with the boundary condition $G(t_f, t_f, \rho) = 1$~\cite{Stadler10,Stadler12}. We obtain
\EQA
b_\alpha (t; t_f, \rho) & = & 
b_\alpha (\rho) \left [1 - \frac{b_\alpha (\rho) + d_\alpha (\rho)}{b_\alpha (\rho) + d_\alpha (\rho) \exp[(b_\alpha (\rho) + d_\alpha (\rho))(t_f-t)]} \right].
\EEA
The observed summary statistics $I_\alpha (t; t_f, \rho)$, $E_\alpha  (t; t_f, \rho)$, and $T_\alpha (t; t_f, \rho)$ follow the dynamical equations 
\EQA
\dot I_\alpha (t; t_f, \rho) & = & b_\alpha (t, t_f, \rho)  \, L_\alpha (t; t_f, \rho)
\nonumber 
\\ 
\dot E_\alpha (t; t_f, \rho) & = & d_\alpha (\rho)  L_\alpha (t; \rho) = \rho d_\alpha L_\alpha (t), 
\nonumber 
\\ 
\dot T_\alpha (t; t_f,\rho) & = & L_\alpha (t; t_f,\rho), 
\nonumber 
\\
\dot L_\alpha (t; t_f,\rho) & = & b_\alpha (t, t_f,\rho) \, L_\alpha (t; t_f,\rho) - d_\alpha\,(\rho)  L_\alpha (t; \rho). 
\label{dot_f}
\EEA
These equations can be integrated analytically. We find
\EQA
L_\alpha (t; t_f,\rho) & = &
C_\alpha (\rho) \, r_\alpha \, \exp (r_\alpha t) \, 
\frac{1 - \exp [- (b_\alpha (\rho) + d_\alpha (\rho)) (t_f - t)] }{1 + \frac{b_\alpha }{ d_\alpha }  \exp[ - (b_\alpha (\rho)+ d_\alpha (\rho)) (t_f - t)]}
\nonumber \\
& = & C_\alpha (\rho) \, r_\alpha \, \exp (r_\alpha t) \, [1 - O(\exp [- (b_\alpha (\rho) + d_\alpha (\rho)) (t_f - t)] ) ] ,
\nonumber \\
I_\alpha (t; t_f, \rho) & = & C_\alpha (\rho) \, b_\alpha (\rho) \, \exp (r_\alpha t) \, \frac{1+\frac{d_\alpha(\rho)}{b_\alpha(\rho)} \exp[-(b_\alpha(\rho)+d_\alpha(\rho)) (t_f-t)]}{1+ \frac{b_\alpha(\rho)}{d_\alpha(\rho)} \exp[-(b_\alpha(\rho)+d_\alpha(\rho)) (t_f-t)]}
\nonumber
\\
&=&C_\alpha (\rho) \, b_\alpha (\rho) \, \exp (r_\alpha t) \, [1 - O(\exp [- (b_\alpha (\rho) + d_\alpha (\rho)) (t_f - t)] ) ] ,
\nonumber \\ 
E_\alpha (t; t_f,\rho) & = & C_\alpha (\rho) \, d_\alpha (\rho) \, \exp(r_\alpha t) ,
\nonumber 
\\ 
T_\alpha (t; t_f,\rho) 
&=& C_\alpha (\rho) \,  \exp (r_\alpha t) \big [ 1 - O \big ( \exp[- (b_\alpha(\rho) + d_\alpha(\rho))(t_f - t)] \big ) \big ] .
\label{bd}
\EEA

The full solution for $T_\alpha (t;t_f,\rho)$ is given in terms of hypergeometric functions. Expanding this solution about $t_f$ shows that termination of sampling does not affect the dynamics (\ref{exp}) and (\ref{exp_rho}) for the bulk of the observation period, but it substantially depletes the observed summary statistics $T_\alpha (t;t_f,\rho)$ and $I_\alpha (t;t_f,\rho)$ over a boundary interval of length $1/(b_\alpha + d_\alpha)$ before the termination point $t_f$. Because there is no boundary depletion for $E_\alpha (t, t_f)$,  the relative amplitudes of the observed summary statistics in the boundary interval are modified compared to the bulk period. In Fig.~S3, the solutions (\ref{exp_rho}) and (\ref{bd}) are plotted and compared to numerical simulations.

\paragraph{Data analysis of Ebola clades.} 
The tree-averaged data of Fig.~2a follow an exponential pattern for most of the observation period; the exponential behavior can also be inferred from individual high-scoring trees (Fig.~S2). Deviations from exponential growth for early times are caused by the stochasticity of birth and death events at low $L_\alpha (t)$. Deviations towards the end of the sampling period can in part be explained by the boundary sampling effects discussed above; the relative amplitudes of the summary statistics change in qualitative agreement with equations (\ref{bd}). In addition, the coherent increase of $E_2 (t)$ and $E_3 (t)$ suggests an increase of the sampling density towards the end of the period. Variations in sampling density can be attributed to differences in reporting, incubation, and continued infectivity after sampling. These factors are only partially accounted for in our and in previous models of the sampling process~\cite{Stadler12}. However, the tree data admit a self-consistent joint fit of the form (\ref{exp_rho}) with $\rho = 0.7$~\cite{Gire14}, which is consistent with homogeneous sampling, over the bulk observation period. From this fit, we estimate the summary statistics $E^o_\alpha$, $I^o_\alpha$, and $T^o_\alpha$ at the end of the period.

\paragraph{Bayesian statistics.}
The summary data $I_\alpha$, $E_\alpha$ and $T_\alpha$ determine the likelihood of a clade-specific, stationary birth-death process as a function of its parameters $b_\alpha$, $d_\alpha$. Given a prior distribution $ P_0  (b_1, \dots, b_k, d_1, \dots, d_k)$ and data from a full genealogical tree, the posterior probability distribution takes the simple form 
\EQA
Q  (b_1, \dots, b_k, d_1, \dots, d_k)  &=&   P_0  (b_1, \dots, b_k, d_1, \dots, d_k) \times 
\nonumber  \\ 
&  & \exp  \sum_{\alpha = 1}^k  \big [ - (b_\alpha + d_\alpha) T_\alpha  + I_\alpha  \log b_\alpha + E_\alpha \log d_\alpha \big ]
\label{S}
\EEA
see refs.~\cite{Stadler10, Stadler12} for a more detailed discussion of genealogical birth-death models. The resulting global maximum-likelihood values are 
\EQ
b_\alpha^*  =  \frac{I_\alpha}{ T_\alpha},
\qquad
d_\alpha^*  =  \frac{E_\alpha}{ T_\alpha}, 
\qquad 
r_\alpha^*  =  \frac{I_\alpha - E_\alpha }{ T_\alpha}, 
\label{ml}
\EE 
assuming a flat prior distribution. The parameters (\ref{ml}) reproduce the relative amplitudes in equations (\ref{exp}), which allows for an important consistency check of the data analysis: the observed time dependence of the summary statistics $I_\alpha (t)$, $E_\alpha (t)$ and $T_\alpha (t)$ in a given clade should match the maximum-likelihood growth rate $r_\alpha$ inferred from the node and branch statistics of that clade. Similarly, we can infer the full-population parameters $b_\alpha$, $d_\alpha$ from the summary statistics $I_\alpha^o$, $E_\alpha^o$, and $T_\alpha^o$ of homogeously sampled genealogical trees. The posterior probability distribution takes the form 
\EQA
Q  (b_1, \dots, b_k, d_1, \dots, d_k)  &= &  P_0  (b_1, \dots, b_k, d_1, \dots, d_k) \times 
\nonumber  \\
&& \exp \sum_{\alpha = 1}^k \bigg[ - (b_\alpha (\rho) + d_\alpha (\rho)) T^o_\alpha  + I^o_\alpha  \log b_\alpha (\rho) + E^o_\alpha \log d_\alpha(\rho) 
\nonumber
\\
&& + \log \frac{\rho (b_\alpha +d_\alpha)}{\sqrt{(b_\alpha-d_\alpha)^2+4 \rho b_\alpha d_\alpha}} \bigg] 
\label{S2}
\EEA
with an additional term generated by the Jacobian of the transformation of variables (\ref{transf}); this term turns out to be numerically small for our data. The inferred  maximum-likelihood parameter values for Ebola clades reported in the main text ($\alpha = 2,3$) have error margins below 10\% due to tree reconstruction uncertainties; we also find our results to be robust to the choice of the BEAST demographic prior. Our analysis uses the simplest clade-specific epidemiological model applicable to this data set; this model allows quantification of sampling effects at high sampling density and avoids the risk of overfitting. Our inference method can be extended to temporally inhomogeneous sampling, which will be the subject of a future publication. 

\paragraph{Evolutionary consequences of clade-specific growth.}
The epidemiological dynamics studied in this paper can be extended to a multi-strain Susceptible-Infected-Recovered (SIR) model of the form introduced by Gog and Grenfell~\cite{Gog02}. In a minimal clade-specific model~\cite{Luksza14}, $S(t)$ denotes the number of hosts susceptible to infection, $L_\alpha (t)$ and $E_\alpha (t)$ are the numbers of hosts infected and after termination of infectivity by a strain of clade $\alpha$, respectively (the notation is chosen to emphasize the relation of these quantities to the tree summary statistics introduced above). In the minimal model, the expectation values of these quantities follow the deterministic dynamical equations, 
\EQA
\dot S (t)  & = & - \frac{S (t)}{S_0} \sum_{\alpha=1}^k b_\alpha L_\alpha (t),  
\label{Sdot}
\\
\dot L_\alpha (t) & = & \left ( b_\alpha \frac{S(t)}{S_0} - d_\alpha \right ) L_\alpha (t) \; \equiv \;  r_\alpha (t) L_\alpha (t), 
\label{Ldot}
\\
\dot E_\alpha (t) & = & d_\alpha L_\alpha (t), 
\label{Edot}
\EEA 
where $S_0$ denotes the initial number of susceptible hosts. This model assumes that all strains are antigenically equivalent. 
The relationship between deterministic SIR dynamics and the underlying stochastic processes is discussed in ref.~\cite{kuehnert13}. The minimal model can be used to derive two generic consequences of heterogeneity in growth parameters:
\begin{enumerate}

\item Evolutionary displacement of clades~\cite{Luksza14}. According to equation (\ref{Ldot}), the clade frequencies 
$X_\alpha (t) = L_\alpha (t) / \sum_\alpha L_\alpha (t)$ follow the evolution equation
\EQ
\dot X_\alpha (t) = [r_\alpha (t) - \bar r(t) ] X_\alpha (t) 
\qquad \mbox{with } \bar r(t) = \sum_{\alpha=1}^k r_\alpha (t) X_\alpha (t).
\EE
For the Ebola clades ($\alpha = 2,3$) in the initial growth phase of the epidemic ($S(t) \approx S_0(t)$), we obtain
\EQ
\frac{X_3 (t)}{ X_2(t)}  = \frac{X_3 (t_0) }{X_2(t_0)} \, \exp [s^* (t - t_0)]
\EE
with $t_0$ at the end of June, $X_2(t_0) = 1 - X_3(t_0) \approx 0.5$, and $s^* = 0.055$. 

\item Acceleration of the epidemic. In the initial growth phase, the net growth rate (or population mean fitness) $\bar r (t)$ increases with time as a result of clade frequency shifts, 
\EQ
\dot{\bar r} (t) = \sum_{\alpha=1}^k \big ( r_\alpha (t) - \bar r(t) \big)^2 X_\alpha (t) + O \left ( \frac{S(t)}{S_0} \right ). 
\EE
\end{enumerate}

\bibliographystyle{plos}
\bibliography{ebola}



\newpage 
\begin{figure}[htbp]
\begin{center}
\includegraphics[scale=0.74]{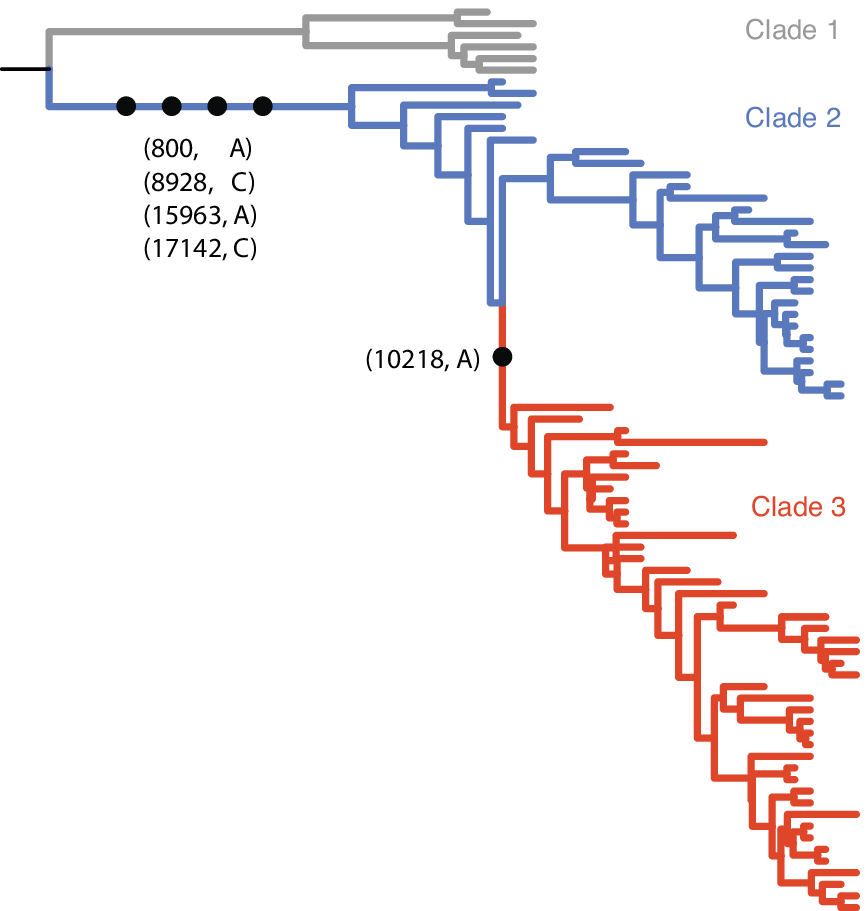}
\caption{{\bf Genealogical tree of the Sierra Leone sequences.}
This tree shows that the Sierra Leone population falls into three genetic clades: the ancestral clade~1 (gray) and the derived clades~2 (blue) and~3 (red). The mutations distinguishing these clades are marked by black dots. Fig.~S1 shows the same tree with sequence annotations.  }
\end{center}
\end{figure}

\begin{figure}[htbp]
\begin{center}
\includegraphics[]{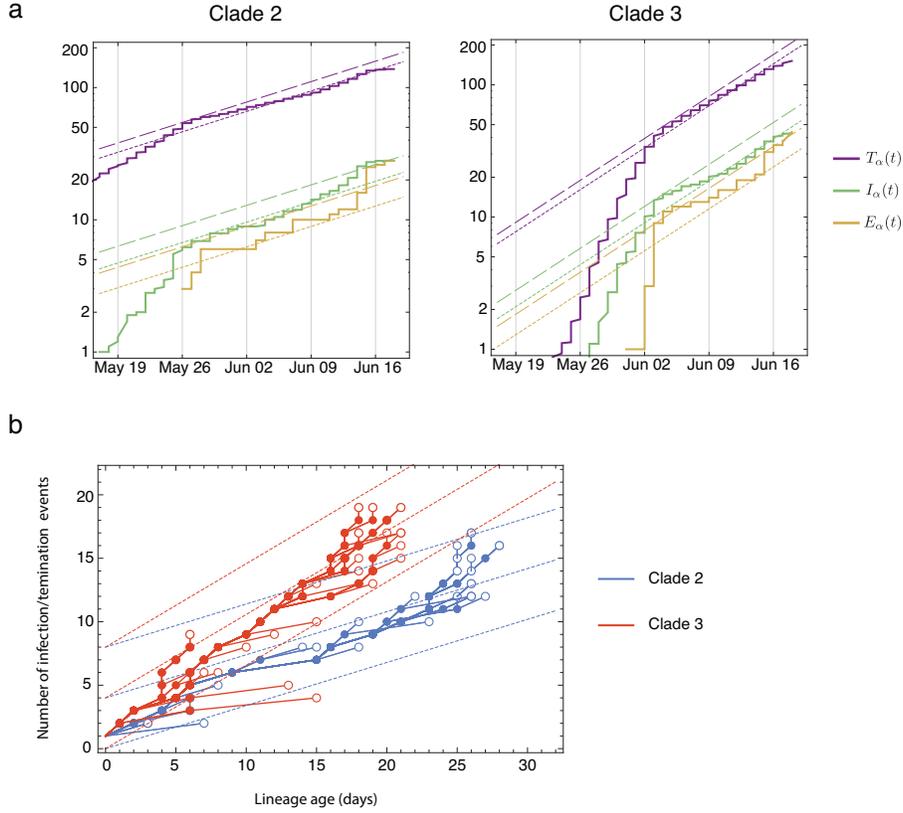}
\caption{{\bf Evolution of the Sierra Leone clades.}
(a)~The number of new infections, $I_\alpha (t)$ (green lines), the number of terminations of infectivity, $E_\alpha (t)$ (yellow lines), and the cumulative infectivity period $T_\alpha (t)$ (purple lines) in clade~$\alpha$ until time $t$ are plotted against $t$ for clade~2 (left) and clade~3 (right). Solid lines: data averaged over 10 high-scoring trees. Short-dashed lines: joint fits to the exponential growth law of equations (\ref{exp_rho}) with the maximum-likelihood parameters $b_\alpha^*$, $d_\alpha^*$ ($\alpha = 2,3$) and a sampling fraction $\rho = 0.7$ (Methods). Long-dashed lines: inferred values of these statistics for a full genealogy, as given by equations (\ref{transf}).  
(b)~The number of infection/termination events up to time $t$ for individual lineages within clade~2 (blue) and clade~3 (red). Infections are shown as filled dots, terminations of infectivity as open circles. Dashed lines with slope $b_\alpha + d_\alpha$ mark the expected rates.}
\end{center}
\end{figure}



\newpage 
\begin{figure}[htbp]
\begin{center}
\includegraphics[width= 0.95 \textwidth]{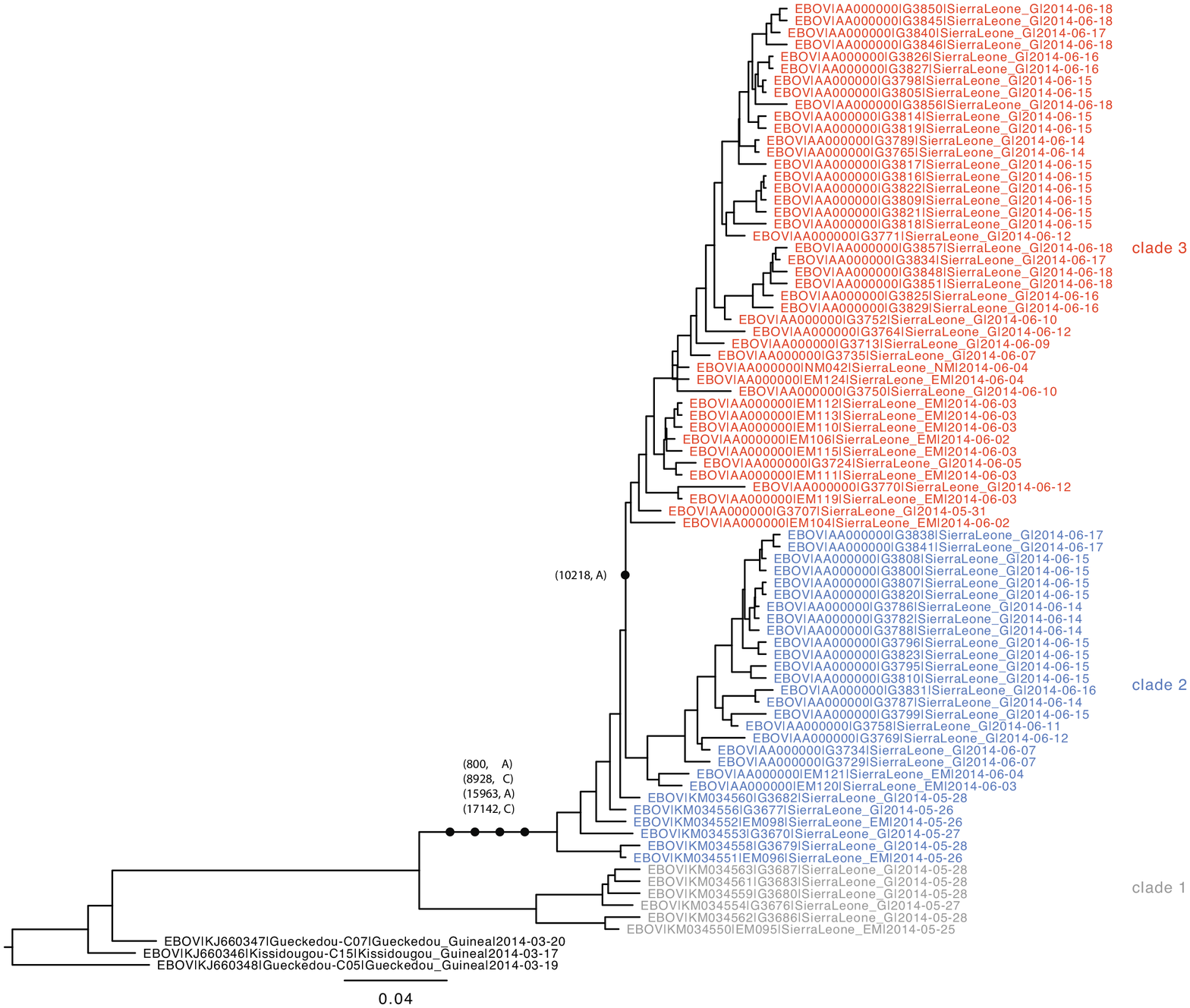}
\end{center}
{\small Figure S1: {\bf Annotated genealogical tree of the Sierra Leone sequences.}
The tree of Fig.~1 is shown with sample names, GenBank  accession numbers and collection dates of all 81 sequences.}
\end{figure}

\newpage 
\begin{figure}[htbp]
\begin{center}
\includegraphics[width= 0.95 \textwidth]{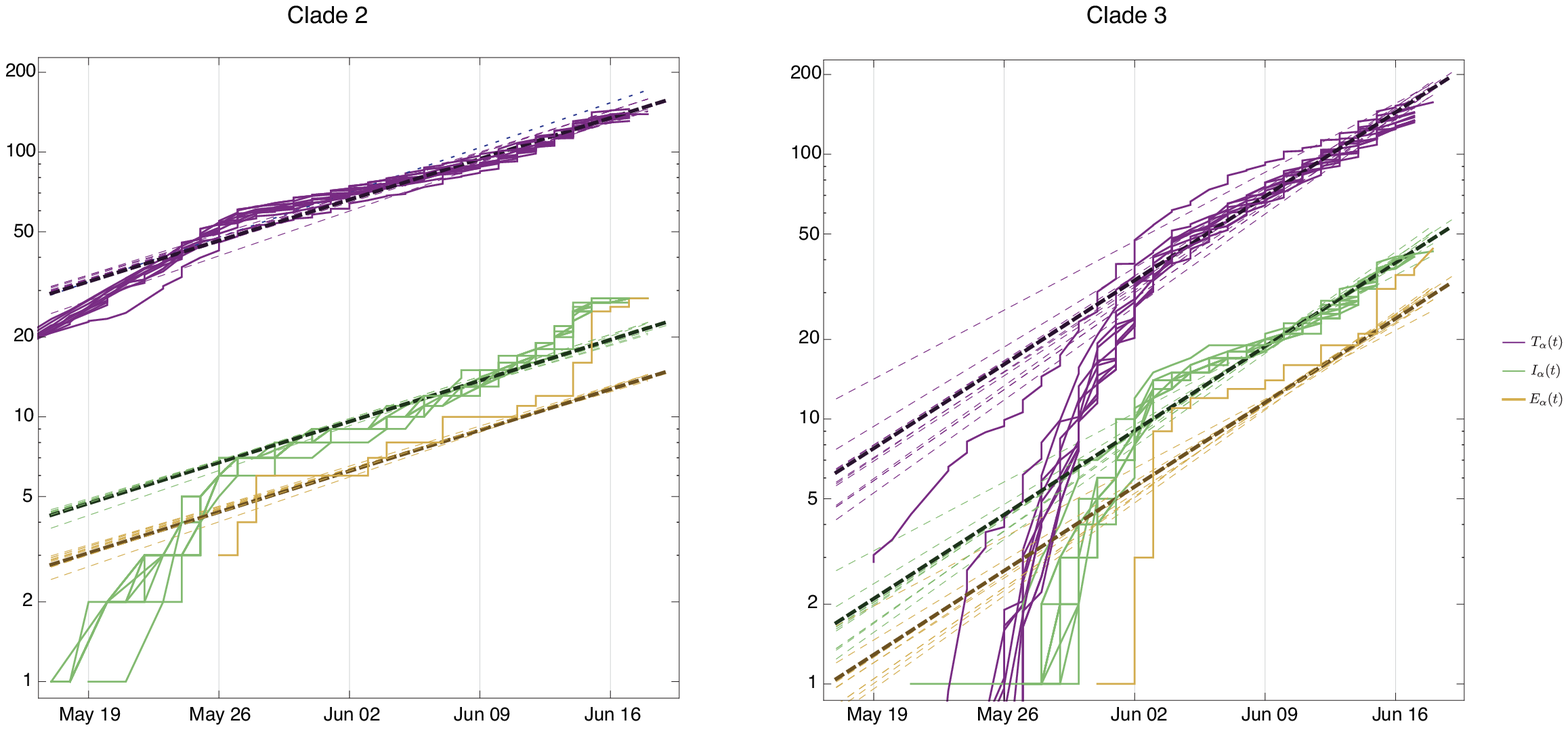}
\end{center}
{\small Figure S2: {\bf Growth variation between inferred genealogies.}
Summary statistics $I_\alpha (t)$ (yellow lines), $E_\alpha (t)$ (green lines), and $T_\alpha (t)$ (purple lines) from 10 high-scoring BEAST trees. Single-tree fits to the exponential growth law (\ref{exp_rho}) (thin dashed lines) are consistent with the clade-specific growth rates (slopes) $r_\alpha$ inferred from tree-averaged data (thick-dashed lines, as in Fig.~2a). 
}
\end{figure}

\newpage 
\begin{figure}[htbp]
\begin{center}
\includegraphics[width= 0.8 \textwidth]{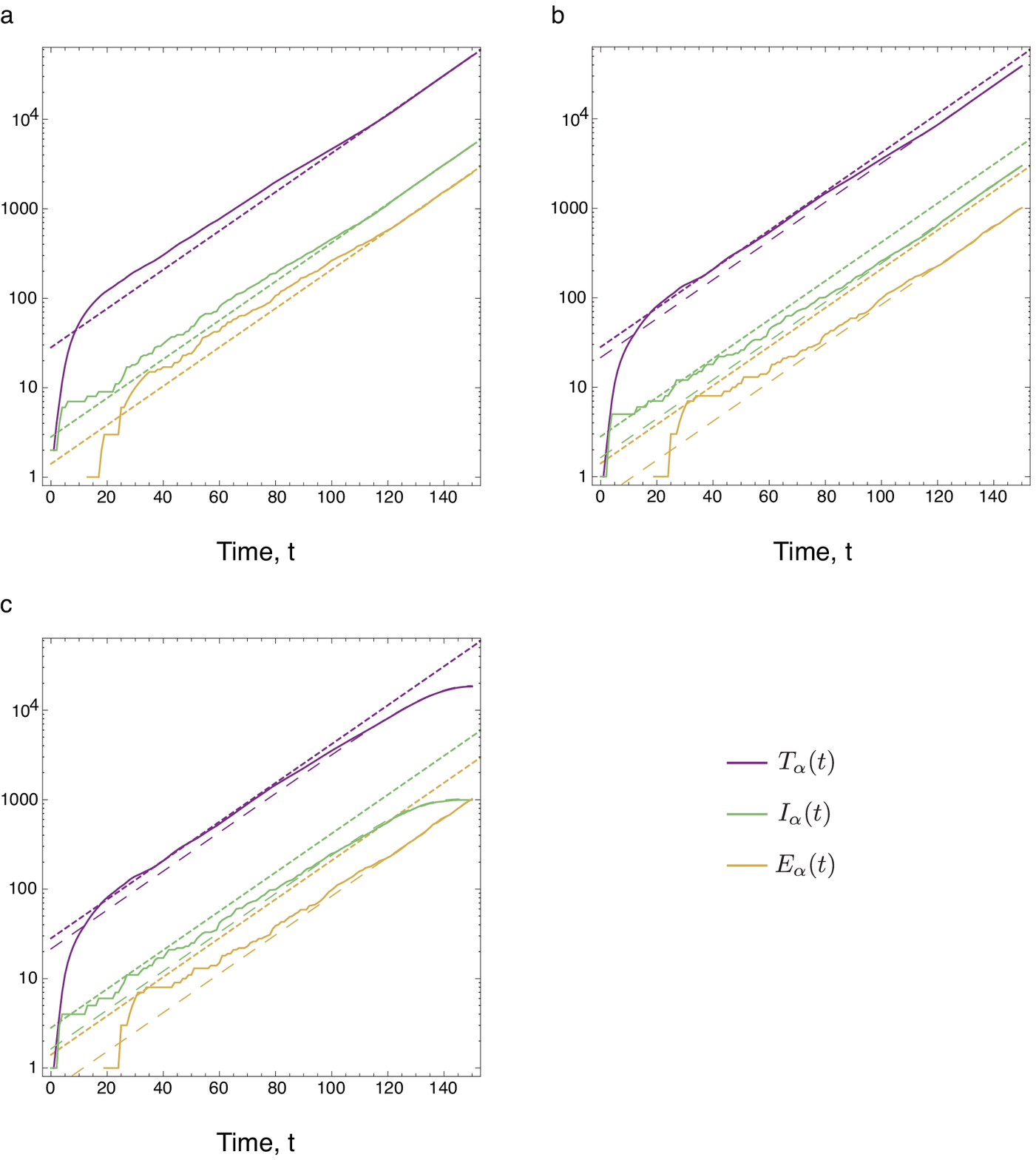}
\end{center}
{\small Figure S3: {\bf Summary statistics of birth-death genealogical trees.}
The time-dependent observables $I(t)$, $E(t)$, and $T(t)$ are shown for a birth-death process with rates $b = 0.10$ and $d = 0.05$ (dashed lines: analytical results, solid lines: simulations).  
(a)~Full tree. The analytical form given by equations (\ref{exp}) is shown by short-dashed lines and is repeated in (b,c) for comparison. 
(b)~Tree with homogeneous sampling (fraction of sampled individuals: $\rho = 0.4$). The analytical form given by equations (\ref{exp_rho}) is shown by long-dashed lines.   Homogeneous sampling modifies the amplitudes of all three observables but leaves their exponential growth at rate $r = b-d$ invariant. 
(c)~Tree with period sampling (fraction of sampled individuals: $\rho = 0.4$, end of sampling period: $t_f = 150$). The analytical form given by equations  (\ref{bd}) is shown by long-dashed lines. Termination of sampling substantially depletes the observables $I(t)$ and $T(t)$ over a boundary interval of length $1/(b+d)$. 
 }
\end{figure}

\end{document}